\documentclass[conference]{IEEEtran}
\usepackage{graphicx,cite,hhline,amsmath,amssymb}

\begin{document}             


\title{\mbox{}\vspace{0.0cm}\\
    On Punctured Pragmatic Space-Time Codes\\
    in Block Fading Channel
    \vspace{0.0cm}}


\author{\authorblockN{Samuele Bandi\authorrefmark{2},
Luca Stabellini,
Andrea Conti and
Velio Tralli}
\authorblockA{\authorrefmark{2}Corresponding author.}
\authorblockA{Authors are with ENDIF, University of Ferrara, via Saragat
1, 44100 Ferrara, Italy  \\ (e-mail: samuele.bandi@gmail.com,lucast17@libero.it,a.conti@ieee.org,vtralli@ing.unife.it)}
}




\maketitle
\begin{abstract}
This paper considers the use of punctured convolutional codes to obtain pragmatic space-time trellis codes over block-fading channel. We show that good performance can be achieved even when puncturation is adopted and that we  can still employ the same Viterbi decoder of the convolutional mother code by using approximated metrics without increasing the complexity of the decoding operations.

\end{abstract}
\IEEEpeerreviewmaketitle
\section{Introduction}
\label{sec:intro}

A relevant result obtained in wireless communications is that the
use of Space-Time Codes (STC) with both multiple transmitting and
receiving antennas can be exploited to mitigate the effect of
fading without sacrificing spectral efficiency. Several research
activities in the last decade has been devoted to find STC able to
improve the performance in terms of achieved diversity and coding
gain \cite{tarok}.  Since the introduction of STC, the problem of
determining the best STC has been always a difficult task,
especially for a large number of transmitting antennas.

In \cite{mcacvt} a new approach to space-time coding called
Pragmatic space-time coding (PSTC) has been devised. The pragmatic
approach consists in using common convolutional codes as STC over a
MIMO channel; it has also been shown that PSTC could be used with
both QPSK and BPSK to achieve maximum diversity and good
performance. The use of convolutional codes as STC allows
decoding by using the same Viterbi decoder with only a proper change in metrics
evaluation.

However, in the case of high-rate codes, $R=k/n$, the number of
operations that the decoder has to perform grows exponentially
with $k$, and the number of paths to be stored increases rapidly.
Puncturation was initially introduced for convolutional codes to
avoid the complexity issue of the Viterbi decoder in case of high
rate codes. Puncturation consists in deleting one or more bits of
a codeword; in this way rate $p/(p+m)$ punctured code can be
obtained by periodically puncturing a low rate $1/n$ convolutional
code, i.e. by erasing $m$ bits for each period of length $p$.

In this paper, following the same aforementioned pragmatic
approach to STC, we will introduced puncturation, using therefore
punctured convolutional codes as ST codes. We will call this
family of codes punctured pragmatic space-time codes (P$^2$-STC).
The use of puncturation in conjunction with STC has also been
investigated in \cite{PuncSTTC}, where the pragmatic approach was
not adopted and only the case of two transmitting antennas was
addressed by designing the codes to preserve full diversity gain
only for short error sequences.

\section{Pragmatic space-time codes:}
In this work we consider P$^2$-STC in a block fading channel (BFC)(see \cite{Chiani} \cite{Elice})
where fading coefficients for each couple of transmitting and receiving antennas are constants
in blocks of $B$ bits and independent block by block. $L$
different blocks are experimented by a codeword. In this
situation, being $N$ the number of transmitting antennas and $M$
the number of receiving antennas, we can generalize the Singleton
bound for BFC to obtain the following diversity bound
\begin{equation}
div \le 1+ \lfloor L N  (1-R) \rfloor \,.
\end{equation}
where $div$ is the maximum achievable diversity degree per receiving antenna.
 
This means that, by puncturing a rate $1/2$ convolutional code in
quasi-static fading channel ($L=1$) with $N=2$,$M=1$, we cannot
achieve $div=2$ since $R>1/2$. For a rate $1/3$ convolutional code  with
$N=3$, $M=1$ and $L=1$, we cannot achieve $div=3$ if $R>1/3$. Our
results prove however, that we can still implement puncturation
and preserve $div=2$ if $1/3<R<2/3$.

\begin{figure}[b]
	\begin{center}
	\centerline{\includegraphics[width=0.95\linewidth,draft=false]{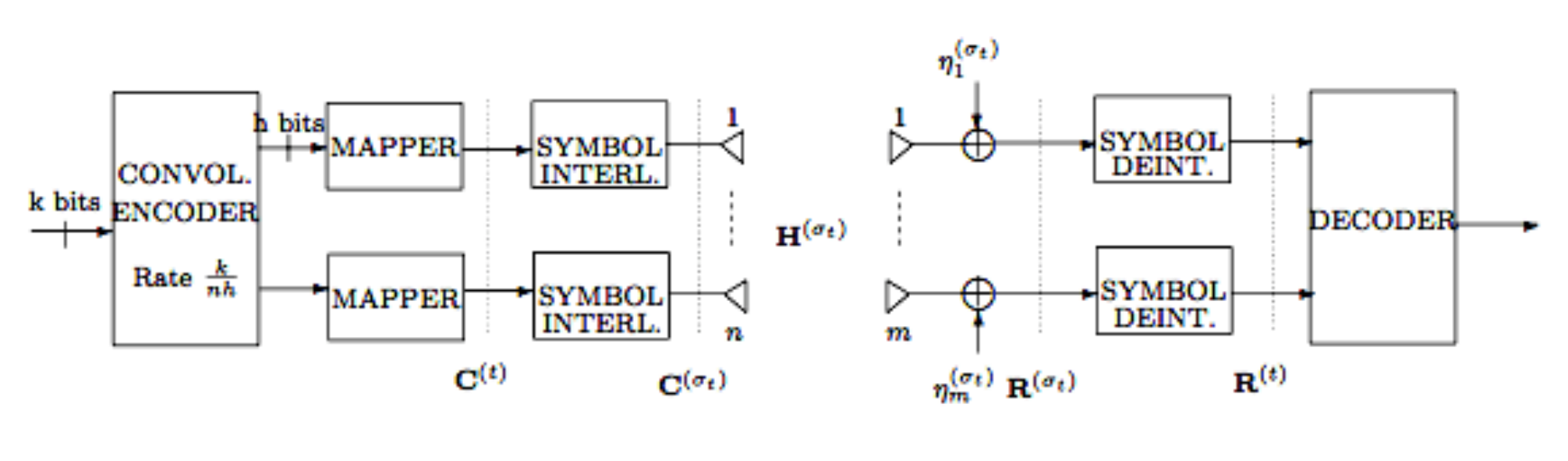}}
	\caption{Block scheme for pragmatic space-time codes (transmitter
side), $4-$states, $1$bps/Hz, BPSK modulation, $n=2$ antennas,
$(5,7)_8$ generators, Decoder stands for Viterbi decoder with proper branch-metrics.}
	\label{schema_blocchi}
	\end{center}
\end{figure}
To describe a system with P$^2$-STC we refer to the block scheme in Fig.\ref{schema_blocchi}. We denote by $\underline{C}^{(t)}=\lbrack{c}^{(t)}_{1}{c}^{(t)}_{2}...{c}^{(t)}_{N} \rbrack^{T}$ a super-symbol, i.e. a  vector of symbols simultaneously transmitted by the $N$ transmitting antennas at the time $t$. A frame is composed by a sequence of $F$ super-symbols (every $T_s$ seconds, N symbols are sent in parallel on the $N$ transmitting antennas). 
If we indicate the complex envelope of the signal received by antenna $s$ at time $t$ with $r_s^{(t)}$ and with $\alpha_{i,j}^{(t)}$ the fading coefficient between antenna $s$ and $i$, at time $t$, we have:
\begin{equation}
r_s^{(t)}=\sum_{i=1}^{n}\alpha_{i,s}^{(t)}c_i^{(t)}\sqrt{E_s}
+\eta_s^{(t)}
\end{equation}
To perform maximum
likelihood decoding, the Viterbi algorithm should estimate the input bit sequence $\hat{\underline{b}}$ in this way:
\begin{equation}
\hat{\underline{b}}={\arg \underset{\underline{b}}{\min}} \sum_t \sum_s |r_s^{(t)}-\sum_{i=1}^N\alpha_{i,s}^{(t)} c_i^{(t)}(\underline{b})\sqrt{E_s}|^2
\end{equation}
where $c_i^{(t)}(\underline{b})$ are the  transmitted symbols corresponding to the path in the trellis labeled by the bit sequence $\underline{b}$, and
\begin{equation}
\Delta M^{(t)}= \sum_s |r_s^{(t)}-\sum_{i=1}^N\alpha_{i,s}^{(t)} c_i^{(t)}\sqrt{E_s}|^2
\end{equation}
are the metric increments.
The fundamental problem is, how to compute these metric increments when
we adopt puncturation. 
Consider for example a rate $1/2$ code used with the puncturation matrix
\begin{equation}
\label{primamat}
P=\left(\begin{array}{cccc}1 & 0 & 1 & 1 \\1 & 1 & 0 & 1\end{array}\right)
\end{equation}
with two transmitting antennas and BPSK modulation.

\begin{figure}[h]	
\begin{center}
\includegraphics[width=0.85\linewidth,draft=false]
{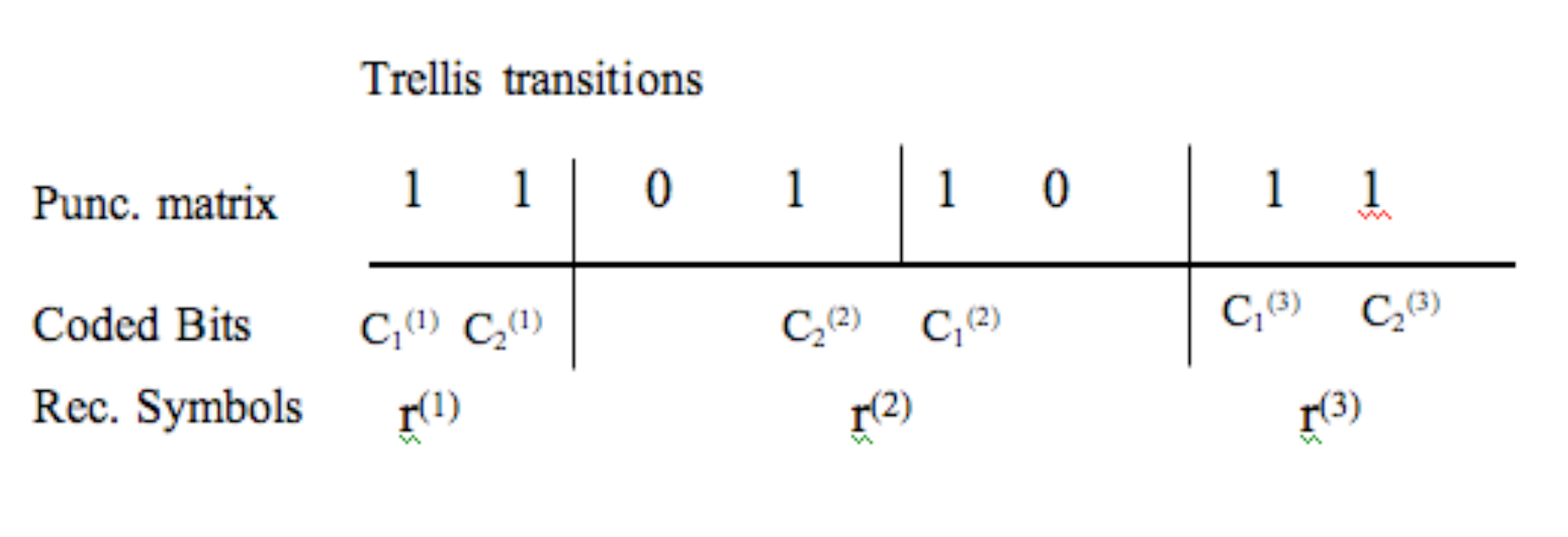}
	\caption{Relationship between received symbols and coded bits for a P$^2$-STC obtained from a R=$1/2$ convolutional code with N=2, M=1}
	\label{figura_1}
	\end{center}
\end{figure}
\noindent

Fig.~\ref{figura_1} shows the relationship between
 the received symbols $r^{(t)}$ and the coded bits.
We can see that the second and third transition of the trellis
diagram are labeled with only one coded bit, i.e., symbol $r^{(2)}$
carries coded bits from two different transitions and therefore we must
use the same received symbol $r^{(2)}$ to compute two transition
metrics on the trellis. Since the coded bit $c_1^{(2)}$ cannot be
recovered from the trellis diagram at instant $t=2$ we can use it
only for the third transition metric. To implement ML
decoding we should join two transitions into one such that no
symbol contains bits from two transitions anymore, as already
pointed out in \cite{PTCM}. This however, would mean that we
could not use a conventional Viterbi decoder of the rate 1/2 code. To exploit the same Viterbi decoder we must actually choose an approximation of ML decoding
that uses approximated metrics. 
In \cite{PTCM} a very similar problem arise in the context of trellis coded modulation. In a similar way we suggest to split the second transition metric in two components that can be used for two different transitions on the trellis diagram. In this way, at the time $t=2$, $ \Delta M^{(t)}$ is approximated with $\Delta \tilde{M}^{(t)}$ as follows:
\begin{eqnarray}
\Delta \tilde{M}^{(t)}= \nonumber \\
1/2\min_{b_2} \sum_s {\vert r_s-[\alpha_{1,s}^{(t)} (2 b_1-1)-\alpha_{2,s}^{(t)} (2 b_2-1)]\sqrt{E_s}\vert }^2   \nonumber \\
+1/2\min_{b1} \sum_s {\vert r_s-[\alpha_{1,s}^{(t)} (2 b_1-1)-\alpha_{2,s}^{(t)} (2 b_2-1)]\sqrt{E_s} \vert }^2
\end{eqnarray}
This metric increment can be easily split on the two transitions at time $t=2$, which we name as left and right transitions supporting transmission at time $t$.

To improve decoding, we also propose
another approximated metric whose computation requires an interaction with the Viterbi algorithm: in other words, with this second approach the left part of the metrics at time $t=2$ can be computed by using the outcomes of the Viterbi algorithm after having processed left transitions at time $t$. 
If the metric computation runs in parallel with the Viterbi algorithm, 
when we know the survivor path at each state of the trellis between left and right transitions, we can compute the metric increments
on the right transition using the information on the survivor path at each departing state $\sigma$. The second approximated metric can be written as
\begin{eqnarray}
\Delta \tilde{M}^{(t)} = \nonumber \\
(1-\beta)\min_{b_2} \sum_s {\vert r_s-[\alpha_{1,s}^{(t)} (2 b_1-1)-\alpha_{2,s}^{(t)} (2 b_2-1)]\sqrt{E_s}\vert }^2 \nonumber \\
+\beta \sum_s {\vert r_s-[\alpha_{1,s}^{(t)} (2 b_1(\sigma)-1)-\alpha_{2,s}^{(t)} (2 b_2-1)]\sqrt{E_s} \vert }^2
\end{eqnarray}
where an additional parameter $\beta$ is included for further optimization.


We have compared these two metrics and we had confirmation that the second one performs better, since it exploits the additional information provided by the Viterbi algorithm. In the next section, we will address the way of generalizing this method of evaluating decoding metrics for different codes with BPSK modulation and different puncturation matrices. 

\section{Decoding metric computation}

When the transmitted super-symbol carries coded bits from two different transitions due to puncturing, the approximated metric increment that exploits the information on the survivor path has proved to give the best performance.
However, up to now, it can only be computed for a puncturation pattern similar to that given in (\ref{primamat}).
The next step is to generalize the metric computation for the generic puncturation matrix and for a generic number of transmitting antennas. We will do this in two steps: first we will find a generalization for the case with two transmitting antennas ($N=2$) and  a generic puncturation matrix and then we will remove this hypothesis and consider the case with $N>2$.

\begin{figure}[h]	
\begin{center}
\includegraphics[width=0.85\linewidth,draft=false]
{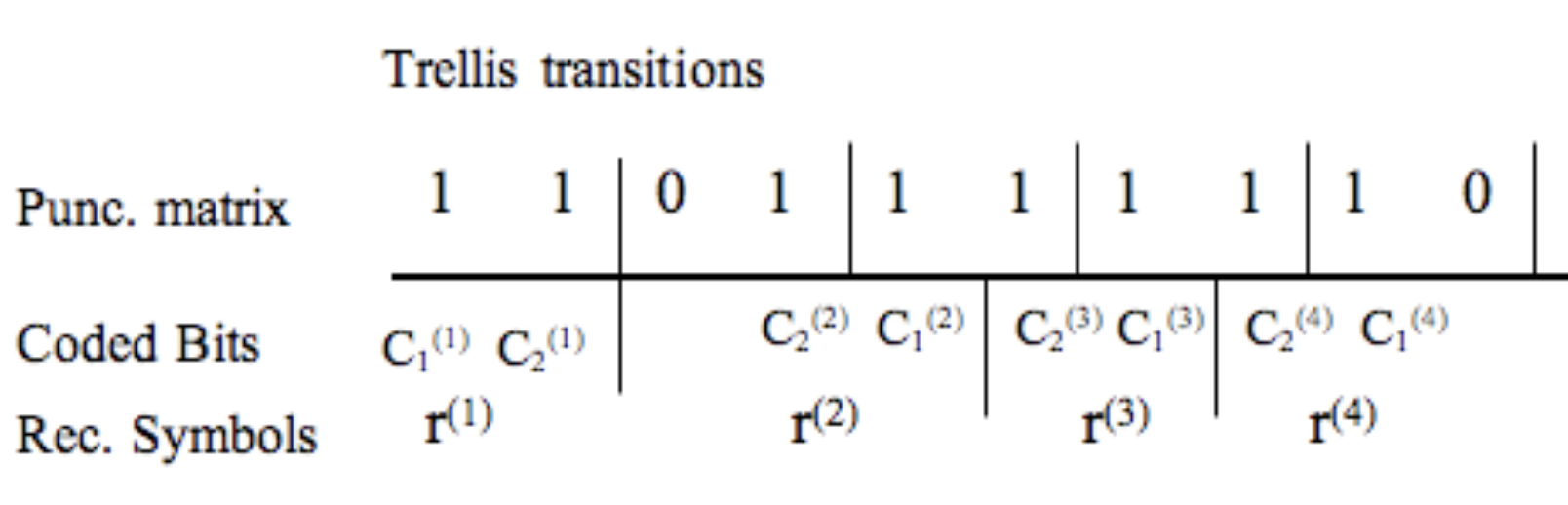}
	\caption{Relationship between received symbols and coded bits of P$^2$-STC obtained from a R=$1/2$ convolutional code with N=2, M=1} and puncturation matrix as in (9).
	\label{transizioni2}
	\end{center}
\end{figure}

We  first considered a two columns puncturation matrix as follows:
 \begin{equation}
 \label{matrice_di_riferimento}
P_1=\left(\begin{array}{cccccccc}1 & 0  & 1 & 1 & 1 & 1 \\1 & 1 & 1 & 1 & 0 & 1\end{array}\right) 
\end{equation}
In this case we have three symbols ($\delta=3$) carrying coded bits belonging to different transitions of the trellis diagram. 
As the decoder computes three metrics using the three received symbols, we must split these three metrics into four trellis transition. 
The relationship between the received symbols $r^{(t)}$ and the coded bits is shown in Fig.\ref{transizioni2}.

Since $r^{(2)}$ carries bits from the second and third transitions on the trellis we will use it to compute the metric increments for the left and right transitions at time $t=2$.  Since also $r^{(3)} $ carries bits from the third transition we will use both $r^{(2)}$ and $r^{(3)}$ to compute the metric for the third transition (which is also the left transition at time $t=3$). On the left metric computed at time $t=3$ the symbol $c_1^{(3)}$ is unknown; therefore this left metric is evaluated with the value of $c_1^{(3)}$ that minimizes it.
On the right metric computed at time $t=2$ the information from the Viterbi algorithm can be exploited to derive the value of 
$c_2^{(2)}$ for the computation of the metric.
%
Each metric increment coming from received symbols carrying bits belonging to two trellis transitions can be split into two components, a left metric $\Delta \tilde{m}_L^{(t)}$ and a right metric $\Delta \tilde{m}_R^{(t)}$: the former will be approximately computed by considering the coded bits of the right transition at time $t$ as unknown, whereas the latter will be computed by considering the coded bits on the left transition at time $t$ as those of the survivor path at the trellis state $\sigma^{(t)}$. That is, for $t=2,3,4$: 
\begin{equation}
 \Delta \tilde{M}^{(t)}= \Delta \tilde{m}_L^{(t)} + \Delta \tilde{m}_R^{(t)} 
 \end{equation}
 where  
\begin{eqnarray}
\Delta \tilde{m}_L^{(t)} &=&(1-\beta)min_{b_2} \sum_s
| r_s^{(t)}-\alpha_{1,s}^{(t)} (2 b_1^{(t)}-1)\nonumber \\ 
&-&\alpha_{2,s}^{(t)} (2 b_2-1)\sqrt{E_s}|^2
\end{eqnarray}
 \begin{eqnarray}
  \Delta \tilde{m}_R^{(t)} &=& \beta \sum_s 
  | r_s^{(t)}-\alpha_{1,s}^{(t)} (2 b_1^{(t)}(\sigma^{(t)})-1)\nonumber \\
  &-&\alpha_{2,s}^{(t)} (2 b_2^{(t)}-1)\sqrt{E_s} |^2
 \end{eqnarray}
In the decoding algorithm for this example, the second transition of the trellis will be labelled with $ \Delta\tilde{m}_L^{(2)}$, the third transition with $\Delta\tilde{m}_R^{(2)}$+$\Delta\tilde{m}_L^{(3)}$, the fourth transition with $\Delta\tilde{m}_R^{(3)}$+$\Delta\tilde{m}_L^{(4)}$ and the last transition with $\Delta\tilde{m}_R^{(4)}$.

We have investigated the optimum value of $\beta$ for different values of $\delta$ and for a rate 1/2 code with puncturation matrices with 2 zeros (as in eq. (9)). We found that the optimum value of $\beta$ is a decreasing function of $\delta$. The reason of this behaviour is that the information held by the survivor path at each state exploited in the right metric will be more and more unreliable as the $\delta$ increases. For $\delta =1$ this value approaches 0.5. 

The generalization of this approach to the $N>2$ case can be easily done if we assume that the puncturation pattern does not allow the transmission of super-symbols carrying coded bits from more than two trellis transition. 
Also in this case each approximated metric increment can be split into a left and right components. These two component may be related to more than one coded bits and the value of $\beta$ should be suitably chosed and optimized for each puncturation pattern, for each value of $N$ and even for each time $t$. However, we can give the following rules of thumb to compute the parameter $\beta$ in the approximated metric increment at time $t$:
 \begin{equation} 
  \begin{array}{ cc}
  \beta-1=\frac{n_L}{n_{tot}},    &  \beta=\frac{n_R}{n_{tot}} 
\end{array}
\end{equation}
 where $n_L$ is the number of not erased coded bits in the label of the left transition at time $t$ and $n_R$ is the number of not erased coded bits in the label of the rigth transition at time $t$. For example, if we refer to matrices in (20), we suggest using $\beta=1/3$ in the approximated metric at time $t=2$.

 \subsection{A simplified metric}
We propose here another possible approach for the generalization of the metric computation, which is sligthly simplified with respect to the one already presented. We name this approximated metric as Type-2 metric to distinguish it from  the former metric named Type-1 metric. With this second approach, if we refer to the example in Fig.\ref{transizioni2} as before, 
we will exploit the information of the survivor path at each state only in the last transition with an erased bit, i.e the right-metric increment is used only at the fifth transition. As we want our decoder to be an approximation of an ML decoder, we have to weight each  component of each metric increment with an appropriate parameter:
the first $\delta$ components (left metrics) and the last component have to be multiplied by parameters whose sum must equal to $\delta$ (equal to 3 in the example). Instead of using $\delta+1$ different parameters we choose to use only two different weights: the first one, $\omega_a$, is used for the left metrics, whereas the second one, $\omega_b$, is used for the last right metric. Once again we include a parameter $\beta$ to balance the two different weights. 
\begin{equation}
\omega_a=(1-\beta)\frac{\delta}{\delta+\beta(1-\delta)}       
\end{equation}
\begin{equation}
     \omega_b=\beta\frac{\delta}{\delta+\beta(1-\delta)}
\end{equation}
With this choice of $\omega_a$ and $\omega_b$, if $\beta=0$ then $\omega_a=1$ and $\omega_b=0$, i.e. we do not trust the right metric using survivor paths information. If $\beta=1$ then $\omega_a=0$ and $\omega_b=1$, i.e., we only exploit the last right metric
If $\beta=0.5$ then $\omega_a=\omega_b$, i.e. we weight the metrics uniformly.
For our decoding rule to be an approximation of the ML decoder we require that:
\begin{equation}
\delta \omega_a+\omega_b=\delta
\end{equation}
which is clearly satisfied by our construction of parameters $\omega_a$ and $\omega_b$.

If we refer to the same example as before, the expression for the approximated metric increments at time $t=2,3,4$ becomes:
  \begin{equation}
\sum_{t=2,3,4} \Delta \tilde{M}^{(t)}= \sum_{t=2,3,4} \Delta \tilde{m}_L^{(t)} + \Delta \tilde{m}_R^{(4)} 
 \end{equation}
 where
 \begin{eqnarray}
  \Delta \tilde{m}_L^{(t)}&= &\omega_a {\underset{b_2}{\min}}
 \sum_s| r_s-\alpha_{1,s}^{(t)} (2 b_1^{(t)}-1)\nonumber \\
  &-&\alpha_{2,s}^{(t)} (2 b_2-1)\sqrt{E_s}|^2
  \end{eqnarray}
  \begin{eqnarray}
\Delta \tilde{m}_R^{(t)} &=&\omega_b \sum_s |r_s-\alpha_{1,s}^{(t)} (2 b_1^{(t)}(\sigma^{(t)})-1)\nonumber \\
&-&\alpha_{2,s}^{(t)} (2 b_2^{(t)}-1)\sqrt{E_s} |^2
\end{eqnarray}

With this approach we have split $\delta$ metric increments corresponding to the $\delta$  symbols received at $t=2,3,4$ into $\delta+1$ metric increments.
As in the previous case the optimum values of $\beta$ is a decreasing function of $\delta$ since the state is more and more untrustworthy as $\delta$ increases. A reasonable value of $\beta$ is between $0.5$ and 1.

In rate-adaptive applications this allows to change simple parameters for metric computation in the Viterbi decoder to decode very different puncturation matrices, keeping the overall trellis structure of the decoder unchanged.

\section{Search for good puncturation matrices}
We address now the issue of finding good puncturation matrices that allow to design the wanted rate $R$ for the P$^2$-STC starting from a convolutional mother code of rate $1/N$ for BPSK modulation (this is the case addressed in this paper: we can easily extend the approach for any pragmatic STC, i.e. as an example by considering a mother code with rate $2/(2N)$ or $1/(2N)$ for QPSK modulation).

As a first step, we investigate the behavior of simple puncturation patterns of $N$ zeros to be used as building blocks for good puncturation matrices with $N$ rows ($N$ is also the number of transmitting antennas). 
We conjecture that the position of a basic pattern inside a puncturation matrix do not affect the performance (or its effect is negligible).

A simple observation will guide our search for good patterns: if the number of super-symbols which carry coded bits of different transitions of the trellis diagram increases, the performance of the decoder get worse because of the increasing for the number of approximated metric increments. On the other hand, if we limit this number to zero by considering a puncturing pattern of $N$ zeros that erases all the bits in a single transition (in this case the entire super-symbol is erased), the performance may be degradated because all the erased bits are concentrated in a single position. 
We consider basic patterns obtained by placing $N$ zeros on one or more subsequent columns (in the latter case $\delta$ is the number of columns). The simulations have proved that putting all zeros in the same column is not the best solution. We made simulations comparing the performance obtained by each different pattern when parameter $\beta$ is fixed to the optimum value for each case.
 
For $N=2$ we obtained as best puncturing patterns the following: 
%
%
\begin{equation} 
\begin{array}{ cc}
P=\left(\begin{array}{cccccccc}
... & 1  & 0 & ... &  \\
... & 0 & 1 & ...& \end{array}\right) &
P=\left(\begin{array}{cccccccc}
... & 0  & 1 & ... &  \\
... & 1 & 0 & ... & \end{array}\right)
\end{array}
\end{equation}
For $N=3$  all the best patterns have  one zero on the first column and two on the second column, e.g.
 \begin{equation}
 \begin{array}{ cc}
P=\left(\begin{array}{cccccccc}
... & 1  & 0 & ... &  \\
... & 1  & 0 & ... &  \\
... & 0 & 1 & ... & 
\end{array}\right) 
  &
P=\left(\begin{array}{cccccccc}
... & 1  & 0 & ... &  \\
 ... & 0 & 1 & ... &  \\
... & 1 & 0 & ... & 
\end{array}\right) 
\end{array}
\end{equation}

As a second step in our search, we try to use this basic patterns to build up more complex puncturation matrices that allow to reach higher code rates.
In the search for good puncturation matrices we put the zeros in a way to satisfy the rate-compatibility rule. In table  \ref{table_2tx} we show  a family of rate-compatible puncturation matrices for the  $N=2$ case with a puncturation matrix with $p=10$. In table \ref{table_3tx}, instead, we addressed the case with three transmitting antennas ($N=3$) and a puncturation matrix with $p=10$. The final rate $R$ of the code will be $p/[(m-p)N]$, where $m$ is the number of basic puncturing patterns of $N$ bits in the matrix.


\begin{table}[h]
\label{table_2tx}
\caption{}
\begin{center}
\begin{tabular}{|r|c|}

\hline
rate & rate-compatible \\ \hline
R=5/9 & 
$\begin{array}{cccccccccc}
1 & 1 & 0 & 1 & 1 & 1 & 1 & 1 & 1 & 1\\
1 & 0  & 1 & 1 & 1 & 1 & 1 & 1 & 1 & 1\\
\end{array} $

\\ \hline
R=5/8 &
$\begin{array}{cccccccccc}
1 & 1  & 0 & 1 & 0 & 1 & 1 & 1 & 1 & 1\\
1 & 0  & 1 & 0 & 1 & 1 & 1 & 1 & 1 & 1\\
\end{array} $
\\ \hline
R=5/7 &
$\begin{array}{cccccccccc}
1 & 1  & 0 & 1 & 0 & 1 & 0 & 1 & 1 & 1\\
1 & 0  & 1 & 0& 1 & 0 & 1 & 1 & 1 & 1\\
\end{array} $ \\ \hline
 R=5/6 &
$\begin{array}{cccccccccc}
1 & 1  & 0 & 1 & 0 & 1 & 0 & 1 & 0 & 1\\
1 & 0  & 1 & 0& 1 & 0 & 1 & 0 & 1 & 1\\
\end{array} $
 \\ \hline
\end{tabular}
\end{center}
\end{table}

\begin{table}[h]
\caption{}
\label{table_3tx}
\begin{center}
\begin{tabular}{|r|c|}

\hline
rate & rate-compatible \\ \hline
R=10/27 & 
$\begin{array}{cccccccccc}
1 & 1 & 0 & 1 & 1 & 1 & 1 & 1 & 1 & 1\\
1 & 1 & 0 & 1 & 1 & 1 & 1 & 1 & 1 & 1\\
1 & 0 & 1& 1 & 1 & 1 & 1 & 1 & 1 & 1\\
\end{array} $

\\ \hline
R=10/24 &
$\begin{array}{cccccccccc}
1  & 0 & 1 & 0 & 1 & 1 & 1 & 1 & 1 &1 \\
1  & 0 & 1 & 0 & 1 & 1 & 1 & 1 & 1 & 1\\
 0  & 1 & 0 & 1 & 1 & 1 & 1 & 1 & 1 & 1\\
\end{array} $
\\ \hline
R=10/21 &
$\begin{array}{cccccccccc}
1  & 0 & 1 & 0 & 1 & 0 & 1 & 1 & 1 & 1\\
1  & 0 & 1 & 0 & 1 & 0 & 1 & 1 &  1&1\\
0  & 1 & 0& 1 & 0 & 1 & 1 & 1 & 1&1\\
\end{array} $ 
\\ \hline
 R=10/18 &
$\begin{array}{cccccccccc}
1  & 0 & 1 & 0 & 1 & 0 & 1 & 0 & 1 &1\\
1  & 0 & 1 & 0 & 1 & 0 & 1 & 0 & 1 &1\\
 0  & 1 & 0& 1 & 0 & 1 & 0 & 1 & 1 &1\\
\end{array} $
 \\ \hline
 R=10/15 &
$\begin{array}{cccccccccc}
1  & 0 & 1 & 0 & 1 & 0 & 1 & 0 & 1& 0\\
 1  & 0 & 1 & 0 & 1 & 0 & 1 & 0 & 1&0\\
0  & 1 & 0& 1 & 0 & 1 & 0 & 1 & 0 & 1\\
\end{array} $
\\ \hline
 
\end{tabular}
\end{center}
\end{table}

\begin{figure}[h]
	
	\begin{center}
	\includegraphics[width=0.95\linewidth,draft=false]
	{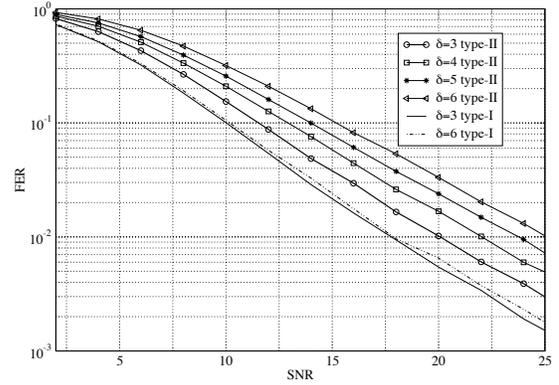}
	\caption{Effects of different basic puncturation patterns of 2 zeros on the performance of the P$^2$-STC scheme with $N=2$. The mother code is(133,171) R=1/2, with a puncturing period $p=10$}
	\label{insensibility}
	\end{center}
\end{figure}

\begin{figure}[h]
	
	\begin{center}
	\centerline{\includegraphics[width=0.95\linewidth,draft=false]{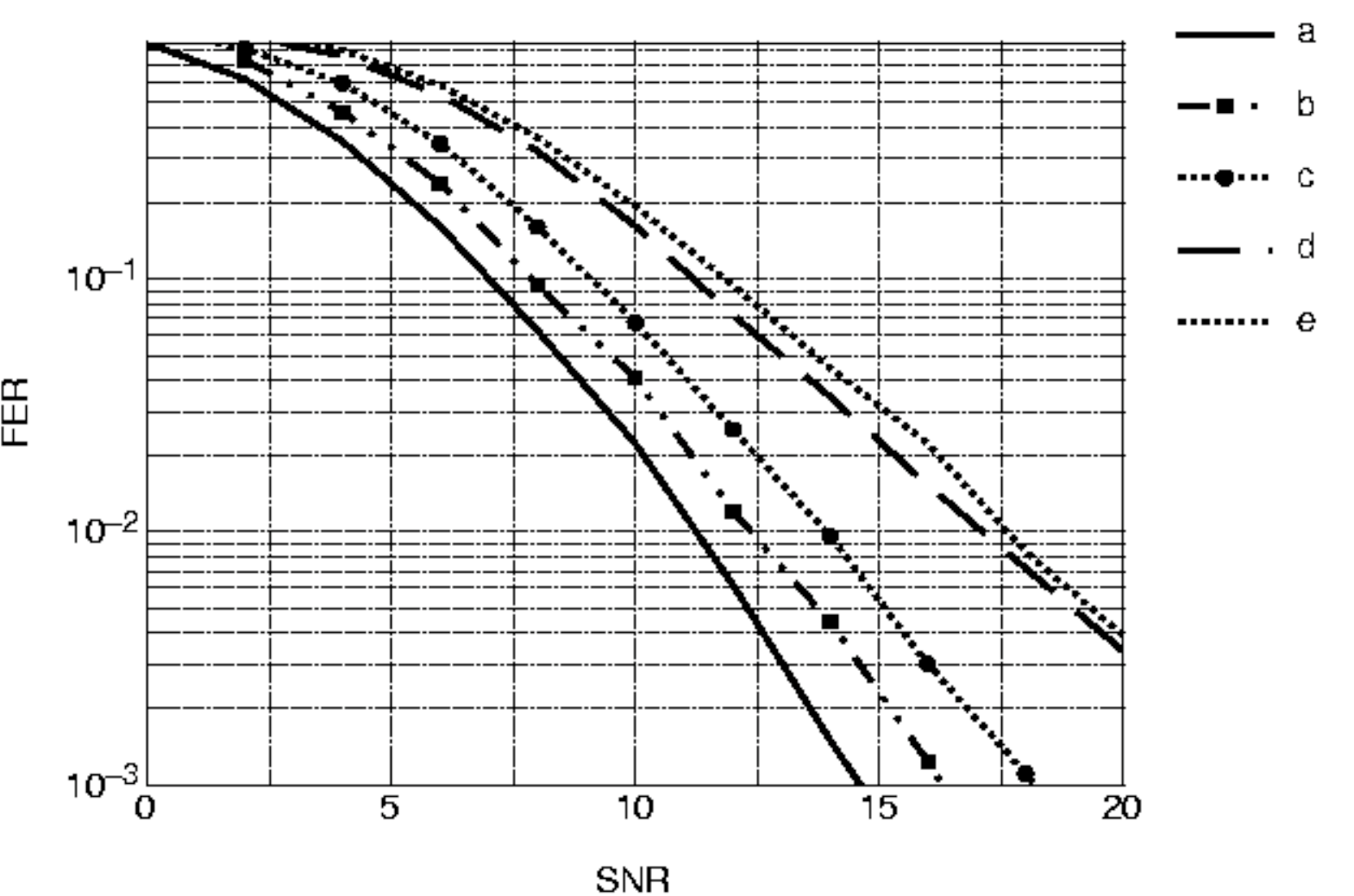}}
	\caption{Performance of P$^2$-STC obtained from a rate 1/3 mother code (133,145,175) and L=1,N=3,M=1. a)R=1/3,  b)R=10/27, c)R=10/24, d)R=10/21, e)R=10/18}
	\label{figura1}
	\end{center}
\end{figure}

\begin{figure}[h]
	
	\begin{center}
	\includegraphics[width=0.95\linewidth,draft=false]
	{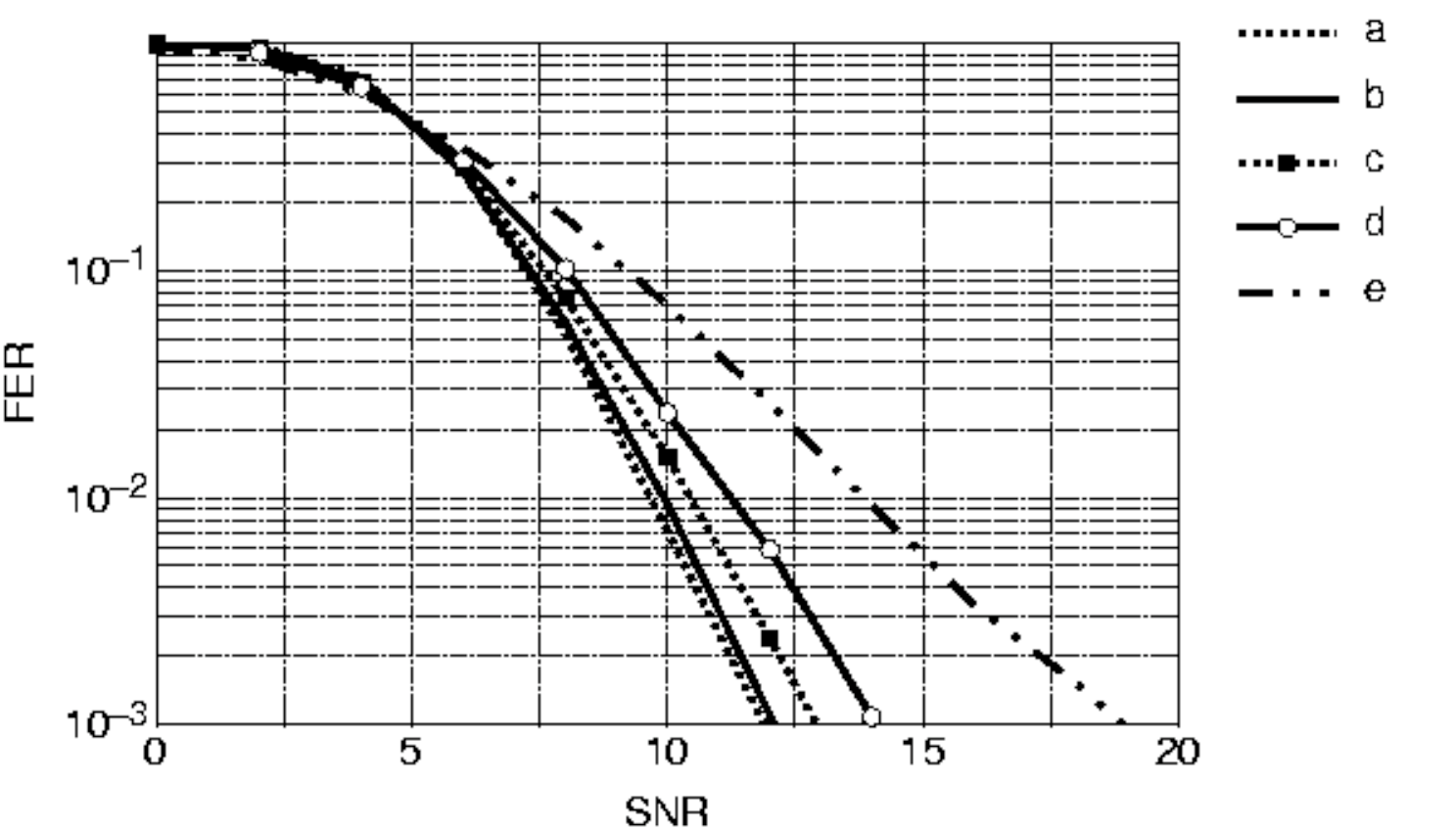}
	\caption{Performance over a BFC of a $R=5/8$ P$^2$-STC obtained from rate 1/2 convolutional  code (133,171) with N=2,M=1. a)L=10,  b)L=8, c)L=6, d)L=4, e)L=2}
	\label{figura2}
	\end{center}
\end{figure}

\section{Numerical results}

The Fig.\ref{insensibility}, shows the sensitivity of the performance of a P$^2$-STC with the same rate to different puncturation patterns of 2 zeros and to the two types of approximated metrics discussed before. We can easily note that Type-I metric is better than Type-II metric as expected, and in general, as already commented in previous sections, the performance degrades as $\delta$ (distance between the 2 zeros) increases. The best case is $\delta=1$, even though for Type-II metric performance changes sligthly.

In Fig.\ref{figura1} we show that it is possible to achieve the maximum diversity allowed on a quasi static channel with a P$^2$-STC with 3 transmitting antennas: for $1/3<R<2/3$, the maximum diversity degree is 2 and can be achieved with all the codes with rate larger than 10/21. It is evident that there is a threashold effect around rate 0.5: if we puncture too much, we lose $1$ degree of diversity.

In Fig.\ref{figura2} we show the performance over a BFC of a rate 5/8 P$^2$-STC obtained from the well-known rate 1/2 mother code (133,171) with a puncturing pattern like the one in (\ref{matrice_di_riferimento}) for varying fading levels per codeword. This code, which can not achieve a diversity degree larger than 1 on a quasi-static channel, is able to
capture a significantly large diversity degree according to the limits given by (1).

\section{Conclusions}
In this paper, we have shown that it is possible to use punctured  convolutional codes to build pragmatic space-codes over a block-fading channel. We have also shown that by an appropriate choice of the metric increments a single Viterbi algorithm on the same trellis diagram can still be used for decoding different rate-compatible codes. We have proposed two different approximated metrics and  explained how to construct rate-compatible puncturation matrices that give good performances. In the results we have shown that with this scheme we can easily obtain codes with rates higher than $1/N$ which can approach or achieve the maximum allowed diversity degree on both quasi-static channel and block-fading channel.

\section {Acknowledgments}
Authors would like to acknowledge Prof. M. Chiani for his helpful discussions. This work has been done within the European Network of Excellence in Wireless Communications (NEWCOM).

\bibliographystyle{IEEEtran}

\begin{thebibliography}{10}
\providecommand{\url}[1]{#1} \csname url@rmstyle\endcsname
\providecommand{\newblock}{\relax}
\providecommand{\bibinfo}[2]{#2}
\providecommand\BIBentrySTDinterwordspacing{\spaceskip=0pt\relax}
\providecommand\BIBentryALTinterwordstretchfactor{4}
\providecommand\BIBentryALTinterwordspacing{\spaceskip=\fontdimen2\font
plus \BIBentryALTinterwordstretchfactor\fontdimen3\font minus
  \fontdimen4\font\relax}
\providecommand\BIBforeignlanguage[2]{{%
\expandafter\ifx\csname l@#1\endcsname\relax
\typeout{** WARNING: IEEEtran.bst: No hyphenation pattern has been}%
\typeout{** loaded for the language `#1'. Using the pattern for}%
\typeout{** the default language instead.}%
\else \language=\csname l@#1\endcsname \fi #2}}

\bibitem{tarok}
Vahid Tarokh, Nambi Seshadri, A. R. Calderbank ``Space-Time Codes for high Data Rate Wireless Communication: Performance Criterion and Code Construction"  \emph{IEEE Transactions on information theory} Volume: 44 no. 2, 2001 Page(s): 744-65 March 1998 

\bibitem{mcacvt}
Chiani, M.; Conti, A.; Tralli, V.; ``A pragmatic approach to space-time coding"  \emph{Communications, 2001. ICC 2001. IEEE International Conference} on, Volume: 9, 2001 Page(s): 2794-2799 vol.9 

\bibitem{PuncSTTC}
Chan-Soo Hwang; Seung Hoon Nam; Jaehak Chung; Byungiang Jeong; ``Design of punctured space-time trellis codes''   \emph{Personal Indoor and Mobile Radio Communications IEEE Proceedings on}, vol.2, pp.
 1698-1702, Sept. 2003.

\bibitem{Chiani}
Chiani, M. ``Error probability for block codes over channel with block interference,''\emph{IEEE Transactions on information theory}   vol.44, Issue 7, pp. 2998-3008, November 1998.
\bibitem{Elice}
McEliece, R.J ; Stark, W.E. ``Channel with block interference'' \emph{IEEE Transactions on information theory}   vol.IT-30, pp. 44-53, Jan. 1984.

\bibitem{PTCM}
Woerz, T;Schweikrt, R. ``Performance of Punctured Pragmatic Codes,'' \emph{Global Telecommunications Conference, '95., IEEE}   vol.1, pp. 13-17, November 1995.
\bibitem{malkamakileib}
 E. Malkamaki, H. Leib,  ``Coded diversity on Block-Fading  Channels" \emph{IEEE Trans. on Information Theory} on, Volume: 45, March 1999 

\end{thebibliography}

\end{document}